\documentstyle[12pt,epsfig]{article}
\def\be{\begin{equation}}
\def\ee{\end{equation}}
\def\bc{\begin{center}}
\def\ec{\end{center}}
\def\bea{\begin{eqnarray}}
\def\eea{\end{eqnarray}}

\def\nn{\nonumber}

\def\gev{{\rm \; GeV}}

\def\smu{\sigma^{\mu}}

\def\grav{\tilde{G}}

\def\mpl{M_{\rm P}}
\def\mt{{\tilde{m}_f}}
\def\mtsq{{\tilde{m}_f^2}}
\def\ov{\overline}

\def\psie{f}
\def\psieb{\overline{f}}
\def\psis{\tilde{G}}
\def\psisb{\overline{\tilde{G}}}

\def\te{\tilde{f}}
\def\tg{\tilde{G}}
\def\dmu{\partial^\mu}
\def\dmd{\partial_\mu}

\def\Lb{\overline{\Lambda}}
\def\Eb{\overline{E}}

\def\elc{\epsilon_{\mu\nu\rho\lambda}}

\catcode`@=11
\def\marginnote#1{}
\newcount\hour
\newcount\minute
\newtoks\amorpm
\hour=\time\divide\hour by60
\minute=\time{\multiply\hour by60 \global\advance\minute by-\hour}
\edef\standardtime{{\ifnum\hour<12 \global\amorpm={am}%
        \else\global\amorpm={pm}\advance\hour by-12 \fi
        \ifnum\hour=0 \hour=12 \fi
        \number\hour:\ifnum\minute<10 0\fi\number\minute\the\amorpm}}
\edef\militarytime{\number\hour:\ifnum\minute<10 0\fi\number\minute}
\def\draftlabel#1{{\@bsphack\if@filesw {\let\thepage\relax
   \xdef\@gtempa{\write\@auxout{\string
      \newlabel{#1}{{\@currentlabel}{\thepage}}}}}\@gtempa
   \if@nobreak \ifvmode\nobreak\fi\fi\fi\@esphack}
        \gdef\@eqnlabel{#1}}
\def\@eqnlabel{}
\def\@vacuum{}
\def\draftmarginnote#1{\marginpar{\raggedright\scriptsize\tt#1}}
\def\draft{\oddsidemargin 0.0truein
        \def\@oddfoot{\sl preliminary draft \hfil
        \rm\thepage\hfil\sl\today\quad\militarytime}
        \let\@evenfoot\@oddfoot \overfullrule 3pt
        \let\label=\draftlabel
        \let\marginnote=\draftmarginnote
   \def\@eqnnum{(\theequation)\rlap{\kern\marginparsep\tt\@eqnlabel}%
\global\let\@eqnlabel\@vacuum}  }
\catcode`@=12
\begin{document}
\begin{titlepage}
\vspace*{-1cm}
hep-th/9709111
\hfill{CERN-TH/97-244}
\\
\phantom{bla}
\hfill{DFPD~97/TH/35}
\\
\vskip 1.0cm
\begin{center}
{\Large\bf On the effective interactions of a light \\ \vskip .2cm
gravitino with matter fermions\footnote{Work supported in part 
by the European Commission TMR Programme ERBFMRX-CT96-0045.}}
\end{center}
\vskip 0.8  cm
\begin{center}
{\large Andrea
Brignole}\footnote{e-mail address: brignole@vxcern.cern.ch}
\\
\vskip .1cm
Theory Division, CERN, CH-1211 Geneva 23, Switzerland
\\
\vskip .2cm
{\large Ferruccio
Feruglio}\footnote{e-mail address: feruglio@padova.infn.it}
\\
\vskip .1cm
Dipartimento di Fisica, Universit\`a di Padova, I-35131 Padua, Italy
\\
\vskip .2cm
and
\\
\vskip .2cm
{\large Fabio
Zwirner}\footnote{e-mail address: zwirner@padova.infn.it}
\\
\vskip .1cm
INFN, Sezione di Padova, I-35131 Padua, Italy
\end{center}
\vskip 0.5cm
\begin{abstract}
\noindent
If the gravitino is light and all the other supersymmetric 
particles are heavy, we can consider the effective theory 
describing the interactions of its goldstino components with
ordinary matter. To discuss the model-dependence of these 
interactions, we take the simple case of spontaneously broken 
supersymmetry and only two chiral superfields, associated with
the goldstino and a massless matter fermion. We derive the 
four-point effective coupling involving two matter fermions and two 
goldstinos, by explicit integration of the heavy spin-0 degrees 
of freedom in the low-energy limit. Surprisingly, our result is 
not equivalent to the usual non-linear realization of supersymmetry, 
where a pair of goldstinos couples to the energy-momentum tensor 
of the matter fields. We solve the puzzle by enlarging the
non-linear realization to include a second independent invariant 
coupling, and we show that there are no other independent couplings
of this type up to this order in the low-energy expansion. We 
conclude by commenting on the interpretation of our results and
on their possible phenomenological implications.
\end{abstract}
\vfill{
CERN-TH/97-244
\newline
\noindent
September 1997}
\end{titlepage}
\setcounter{footnote}{0}
\vskip2truecm
{\bf 1.}
It is quite plausible that the theory of fundamental 
interactions lying beyond the Standard Model has a
spontaneously broken $N=1$ space-time supersymmetry
(for reviews and references, see e.g. \cite{susy}). 
However, the dynamical origin of the energy scales 
controlling supersymmetry breaking is still obscure, 
and different possibilities can be legitimately considered.
In this paper, following the general strategy outlined in 
\cite{bfz3}, we concentrate on the possibility that 
the gravitino mass $m_{3/2}$ is much smaller than 
all the other supersymmetry-breaking mass splittings.
In this case, the $\pm 1/2$ helicity components of the 
gravitino, corresponding to the would-be goldstino $\grav$, 
have effective couplings with the various matter and gauge
superfields much stronger than the gravitational ones. 
Exploiting the supersymmetric version of the equivalence 
theorem \cite{equiv}, in a suitable energy range we can
neglect gravitational interactions and define a 
(non-renormalizable) effective theory with spontaneously 
broken global supersymmetry. 

In this general framework, we analyze the low-energy 
amplitudes involving two goldstinos and two matter fermions. 
According to the low-energy theorems for goldstino interactions 
\cite{dwf}, such amplitudes are controlled by the energy-momentum 
tensor $T_{\mu\nu}$ of the matter system. Indeed, explicit 
non-linear realizations of the supersymmetry algebra have been 
built \cite{nl,nlbis}, and they precisely reproduce the behaviour 
prescribed by the low-energy theorems. In the present note, we 
follow an alternative procedure \cite{bfz3}, starting from a theory 
where supersymmetry is linearly realized, although spontaneously
broken, and the building blocks are all the superfields containing 
the light degrees of freedom. Restricting ourselves to energies 
smaller than the supersymmetry-breaking mass splittings, we 
solve the equations of motion for the heavy superpartners
in the low-energy limit, and derive an effective theory
involving only the goldstino and the light Standard Model 
particles, where supersymmetry is non-linearly realized. 
We finally compare the results obtained via this
explicit procedure with those obtained by direct construction of 
the non-linear lagrangian, on the basis of the transformation 
properties of the goldstino and the matter fields. 

A similar program has already been successfully 
implemented in a number of cases. In the simple case
of a single chiral superfield, the effective low-energy 
four-goldstino coupling was computed \cite{ccdfg}, and
the result can be shown to be physically equivalent to the 
non-linear realization of \cite{nl}, in the sense that 
they give rise to the same on-shell scattering amplitudes. 
More recently, we computed the effective low-energy coupling 
involving two photons and two goldstinos \cite{bfz3}. Our result 
can be shown to be physically equivalent, in the same sense 
as before, to the non-linear realization of \cite{nlbis}, 
where goldstino bilinears couple to the canonical 
energy-momentum tensor of matter and gauge fields.
 
In this paper, we discuss an interesting feature that emerges
when we consider the effective low-energy coupling involving
two goldstinos and two matter fermions. To make the case as clear 
and simple as possible, we consider only one massless left-handed
matter fermion, we turn off gauge interactions and we impose a 
global $U(1)$ symmetry associated with matter conservation\footnote{
With the given fermion content this symmetry is anomalous, but we
can introduce a third chiral superfield, associated with a left-handed 
antimatter fermion $f^c$, that cancels the anomaly without affecting 
any of the following considerations. Also the other assumptions can 
be eventually relaxed, with no impact on our main result.}. In contrast 
with the previous cases, the outcome of our calculation turns out 
to be physically inequivalent to the non-linear realization of 
\cite{nlbis}. To solve the puzzle, we go back to the superfield 
construction of non-linear realizations for goldstinos and matter 
fermions. We show that we can add to the invariant lagrangian, 
associated with the non-linear realization of \cite{nlbis}, a second 
independent invariant, which contributes to the four-fermion 
interaction under consideration. The terms of this additional
invariant containing two goldstinos cannot be expressed in terms of
the energy-momentum tensor of the matter fermion. We also show
that the most general form for the amplitude under consideration
can indeed be parametrized, to this order in the low-energy
expansion, in terms of only two supersymmetric invariants.
After some comments on the interpretation of our results and 
on the open problems, we conclude with some anticipations 
\cite{bfz4} on the possible phenomenological implications.

\vspace{1cm}
{\bf 2.}
As announced in the introduction, we consider an $N=1$ globally 
supersymmetric theory containing only two chiral superfields. One 
of them will describe the goldstino $\tg$ and its complex spin-0
partner $z \equiv (S + i P) / \sqrt{2}$. The other one will describe 
a massless left-handed matter fermion $f$ and its complex spin-0 
partner $\te$. According to the standard formalism \cite{wb}, and 
neglecting for the moment higher-derivative terms, the lagrangian is 
completely specified in terms of a superpotential $w$ and a K\"ahler 
potential $K$. To have spontaneous supersymmetry breaking, and to 
consistently identify $\tg$ with the goldstino, we assume that, at 
the minimum of the scalar potential, 
\be
\langle  F^0  \rangle \ne 0 \, ,
\;\;\;\;\;
\langle  F^{1}  \rangle = 0 \, ,
\label{fterms}
\ee
where $F^0$ and $F^1$ denote the auxiliary fields associated with
the goldstino and with the matter fermion, respectively. It will 
not be restrictive to assume that $\langle z \rangle \! = 0$. We 
shall also assume that $\langle \te \rangle \! = 0$, consistently 
with an unbroken global $U(1)$ symmetry associated with matter 
conservation. 

We proceed by expanding the defining functions of the theory 
around the vacuum, in order to identify the terms contributing 
to the effective four-fermion interaction involving two matter 
fermions and two goldstinos. Without loss of generality, we can 
write:
\be
w=\hat{w}(z)+ \ldots \, , 
\;\;\;\;\;
K=\hat{K}(z,\bar{z})+\tilde{K}(z,\bar{z}) \, |\te|^2
+\ldots \, ,
\label{expan}
\ee
where the dots denote terms that are not relevant for our 
considerations. Taking into account eqs.~(\ref{fterms}) and 
(\ref{expan}), the mass spectrum of the model can be easily
derived from standard formulae \cite{wb}. The goldstino and the 
matter fermion remain massless, whilst all the spin-0 particles 
acquire in general non-vanishing masses, proportional to $\langle 
F^0 \rangle$ and expressed in terms of $w$, $K$ and their 
derivatives, evaluated on the vacuum. Moreover, even in the 
presence of non-renormalizable interactions, the expansion of the 
lagrangian in (canonically normalized) component fields can be 
rearranged in such a way that all the terms relevant for our 
calculation are expressed in terms of the mass parameters $(m_S^2,
m_P^2,\mtsq)$, associated with the spin-0 partners of the goldstino 
and of the matter fermion, and the scale $F$ of supersymmetry 
breaking, without explicit reference to $w$ and $K$:
\bea
{\cal L} 
& = &
 {1\over 2} \left[ (\dmu S) (\dmd S) - m_S^2 S^2 \right]        
+{1\over 2} \left[ (\dmu P) (\dmd P) - m_P^2 P^2 \right]      
+(\dmu \te)^* (\dmd\te)- \mtsq |\te|^2 
\nn \\
& + &
i \psisb \ov{\sigma}^{\mu} \partial_{\mu} \psis 
+
i \psieb \ov{\sigma}^{\mu} \partial_{\mu} \psie 
- 
{1\over 2\sqrt{2} F} [ (m_S^2 S + i m_P^2 P) \psis 
\psis + (m_S^2 S - i m_P^2 P) \psisb \psisb ]
\nn \\
&-&{\mtsq \over F} (\te^* \, \psis \psie +\te \, \psisb \psieb) 
-{\mtsq \over F^2} \, \psis \psie \, \psisb \psieb + \ldots \, .
\label{lag}
\eea
In eq.~(\ref{lag}), we have used two-component spinors with the 
conventions of \cite{bfz3}. The parameter $F \equiv \, < \! 
\ov{w}_{\ov{z}} (K_{\ov{z} z})^{-1/2} \! >$ (lower indices denote
derivatives) defines the supersymmetry-breaking scale and has the 
dimension of a mass squared. For simplicity, we have assumed $F$ 
to be real. We recall that, in our flat space-time, $F$ is linked 
to the gravitino mass $m_{3/2}$ by the universal relation $F^2 = 
3 \, m_{3/2}^2 \, \mpl^2$, where $\mpl \equiv (8 \pi G_N)^{-1/2} 
\simeq  2.4 \times 10^{18} \gev$ is the Planck mass. Finally, the 
dots in eq.~(\ref{lag}) stand for terms that do not contribute to 
the four-fermion amplitudes of interest\footnote{There are interaction 
terms proportional to $<\tilde{K}_z>$ and $<\tilde{K}_{\ov{z}}>$, 
not explicitly listed here, that are in principle relevant. An 
explicit computation shows that their total contribution vanishes. 
This is in agreement with the possibility of choosing normal 
coordinates \cite{grk}, where such terms are absent.}.

Starting from the lagrangian of eq.~(\ref{lag}), we take the limit
of a heavy spin-0 spectrum, with $(m_S,m_P,\mt)$ much larger than 
the typical energy of the scattering processes we would like to study. 
In this case, we can build an effective lagrangian for the light 
fields by integrating out the heavy states. As discussed in detail
in \cite{bfz3}, the crucial property of such an effective lagrangian 
will be its dependence on the supersymmetry-breaking scale $F$ only, 
without any further reference to the supersymmetry-breaking masses 
$(m_S,m_P,\mt)$. This property is the result of subtle cancellations 
among
\begin{figure}[ht]
\vspace{-0.01cm}
\begin{center}
\epsfig{figure=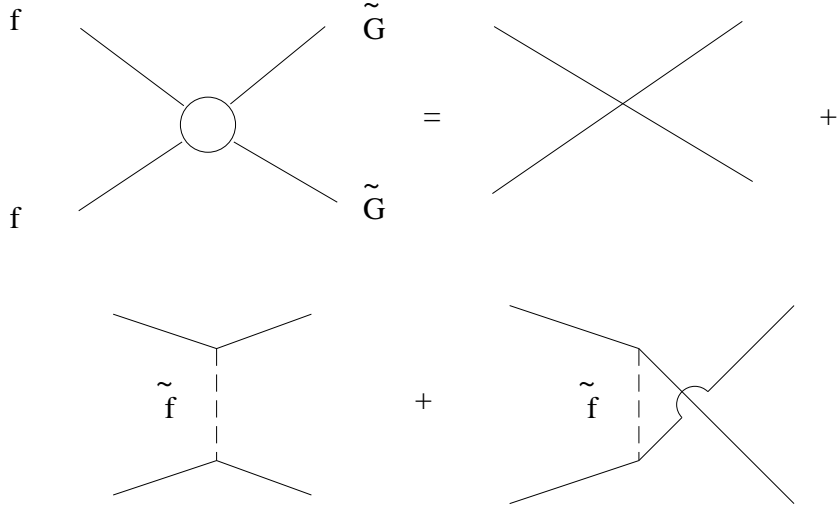,height=11cm,angle=-90}
\end{center}
\caption{{\it Diagrammatic origin of the four-fermion operator of 
eq.~(\ref{leff}).}}
\label{origin}
\end{figure}
the different diagrams shown in fig.~\ref{origin}, corresponding
to the contact term in the last line of eq.~(\ref{lag}) and to 
$\te$ exchange, and agrees with general results [3--6] concerning 
low-energy goldstino interactions. Focussing only on the terms 
relevant for our calculation, we obtain a local interaction term 
involving two matter fermions and two goldstinos, of the form 
\be
{{\cal L}}_{eff} = 
{1 \over F^2} [ \partial_{\mu} ( \psie \psis )]
[ \partial^{\mu} ( \psieb \psisb ) ] + \ldots \, .
\label{leff}
\ee

An alternative derivation of ${{\cal L}}_{eff} $ is possible,
following a technique introduced in \cite{rocek}. Denoting
by $\phi$ and $\phi_f$ the superfields associated with the 
goldstino and with the matter fermion, respectively, we
can impose the supersymmetric constraints $\phi^2 = 0$ a
and $\phi \, \phi_f = 0$, and solve for the fermionic 
components imposing eq.~(\ref{fterms}). The result  
coincides with eq.~(\ref{leff}).

\vspace{1cm}
{\bf 3.}
Could we have derived the effective interaction of eq.~(\ref{leff})
from the non-linear realizations of the supersymmetry algebra
that have been proposed up to now in the literature? 
To address
this question, we recall that the non-linear realization of 
\cite{nl,nlbis} prescribes an effective interaction of the form
\be
\label{leffbis}
{{\cal L}}_{eff}' = 
{i \over 2 F^2} [ \psis \sigma^{\mu} \partial^{\nu} \psisb
- ( \partial^{\nu} \psis ) \sigma^{\mu} \psisb ] \; T_{\nu \mu} 
+ \ldots \, ,
\ee
where $T_{\nu \mu}$ is the canonical energy-momentum tensor
of the matter fermions,
\be
T_{\nu \mu} = i \psieb \ov{\sigma}_{\nu} \partial_{\mu}
\psie + \ldots \, ,
\label{timunu}
\ee
and the dots stand for terms that do not contribute to the on-shell
scattering amplitudes with two matter fermions and two goldstinos. 
Combining (\ref{leffbis}) with (\ref{timunu}), we obtain:
\be
\label{leffter}
{{\cal L}}_{eff}' = 
- {1 \over F^2}  (\psis \sigma^{\mu} \partial^{\nu} \psisb) 
(\psieb \ov{\sigma}_{\nu} \partial_{\mu} \psie) + \ldots \, ,
\ee
which looks very different from (\ref{leff}). 

To check that (\ref{leff}) and (\ref{leffter}) are really 
inequivalent, we concentrate on the scattering amplitudes for the 
process\footnote{This process was already considered by Fayet 
\cite{fayet}, who gave the correct scaling law of the cross-section 
with respect to the gravitino mass and to the centre-of-mass energy 
in the low-energy limit.}
\be
f \; \ov{f} \longrightarrow \grav \grav \, ,
\ee
even if $f \grav \to f \grav$, $\ov{f} \grav \to \ov{f} \grav$
or $\grav \grav \to f \ov{f}$ would be equally good processes 
for this purpose. We denote by $(p_1,p_2,q_1,q_2)$ the 
four-momenta of the incoming fermion and antifermion and of 
the two outgoing goldstinos, respectively. Notice that the
only helicity configurations that can contribute to the 
process are, in the same order of the momenta and in obvious 
notation, $(L,R,L,R)$ and $(L,R,R,L)$.

On the one hand, from the effective lagrangian of eq.~(\ref{leff}) 
we obtain the amplitudes:
\be
\label{amp}
a(L,R,L,R) = -{(1 +\cos \theta)^2 s^2 \over 4 F^2} \, ,
\;\;\;\;
a(L,R,R,L) =  {(1 -\cos \theta)^2 s^2 \over 4 F^2} \, ,
\ee
where $\sqrt{s}$ and $\theta$ are the total energy and the 
scattering angle in the centre-of-mass frame, leading to a 
total cross-section
\be
\label{sigma}
\sigma (f {\bar f} \to \grav \grav) = 
{s^3 \over 80 \pi F^4} \, .
\ee

On the other hand, from the effective lagrangian of eq.~(\ref{leffter}) 
we obtain:
\be
\label{ampter}
a'(L,R,L,R) = {\sin^2 \theta s^2 \over 4 F^2} \, ,
\;\;\;\;
a'(L,R,R,L) = - {\sin^2 \theta s^2 \over 4 F^2} \, ,
\ee
leading to a total cross-section
\be
\label{sigmater}
\sigma' (f {\bar f} \to \grav \grav) = 
{s^3 \over 480 \pi F^4} \, .
\ee

We conclude that the two effective interactions (\ref{leff})
and (\ref{leffter}) lead to the same energy dependence, but 
to different angular dependences and total cross-sections.
Surprisingly, the two approaches seem to give physically 
different results\footnote{The above results can be easily extended 
to Dirac fermions, upon introduction of a second Weyl spinor $f^c$. 
For example, the total unpolarized cross section $\sigma(e^+ e^- 
\to {\tilde G} {\tilde G})$ inferred from (\ref{sigma}) would read 
$s^3/(160 \pi F^4)$ and that from (\ref{sigmater}) $s^3/(960 \pi F^4)$.
Incidentally, we observe that both results 
are in disagreement with a previous computation \cite{eegg}, which 
found $\sigma (\grav \grav \to e^+ e^-) = s^3 / (20 \pi F^4)$, 
corresponding to $\sigma (e^+ e^- \to \grav \grav) = s^3 / (40 
\pi F^4)$.}.

\vspace{1cm}
{\bf 4.}
To understand the origin of the discrepancy, we go back to the 
superfield construction of the non-linear realization of \cite{nlbis}.
This is given in terms of the superfield
\be
\label{bigl}
\Lambda_{\alpha} ( x,\theta,\ov{\theta} ) 
\equiv \exp (\theta Q + \ov{\theta} \ov{Q} ) \, 
\grav_{\alpha} (x) 
=  \grav_{\alpha} + \sqrt{2} F \theta_{\alpha} +
{i \over \sqrt{2} F} ( \grav \sigma^{\mu} \ov{\theta} 
- \theta \sigma^{\mu} \ov{\grav} ) \partial_{\mu} 
\grav_{\alpha} + \ldots  \, ,
\ee
whose lowest component is the goldstino $\grav$, and a superfield
\be
\label{bige}
E_{\alpha} ( x,\theta,\ov{\theta} )  \equiv 
\exp (\theta Q + \ov{\theta} \ov{Q} ) \, f_{\alpha} (x)
=  f_{\alpha} + {i \over \sqrt{2} F} ( \grav \sigma^{\mu} 
\ov{\theta} - \theta \sigma^{\mu} \ov{\grav} ) \partial_{\mu} 
f_{\alpha} + \ldots  \, ,
\ee
whose lowest component is the matter fermion $f$. In the 
simple case under consideration, the non-linear realization of 
\cite{nlbis} can be introduced via the supersymmetric lagrangian
\be
\label{invone}
{1 \over 4 F^4} \int d^4 \theta \; \Lambda^2 \ov{\Lambda}^2
\; i \ov{E} \ov{\sigma}^{\mu} \partial_{\mu} E \, , 
\ee  
which leads precisely to the result of eq.~(\ref{leffter}),
as can be easily verified by an explicit computation.

The crucial question is now the following: are there other
independent invariants, besides (\ref{invone}), that can
contribute to the effective interaction under consideration?
The answer is positive, since a second invariant can be
constructed:
\be
\label{invtwo}
{\alpha \over F^2} \int d^4 \theta \; \Lambda E \; 
\ov{\Lambda} \ov{E} \, , 
\ee
where $\alpha$ is an arbitrary dimensionless coefficient. 
The new invariant (\ref{invtwo}) gives, among other things, 
the following contribution to the four-fermion effective 
interaction under consideration:
\be
\delta {\cal L}_{eff}' = 
{\alpha \over 4 F^2} (\psis \sigma^{\mu} \partial^{\nu} 
\psieb) (\psisb \ov{\sigma}_{\nu} \partial_{\mu} \psie)
+ \ldots \, ,
\label{deltal}
\ee
where the dots stand for terms not contributing to the on-shell process
under consideration. From the contact interaction displayed in eq.
(\ref{deltal}) we obtain the following non-vanishing amplitudes:
\be
\delta a'(L,R,L,R) = \alpha {(1 +\cos \theta) s^2 \over 8 F^2} \, ,
\;\;\;\;
\delta a'(L,R,R,L) = -\alpha {(1 -\cos \theta) s^2 \over 8 F^2} \, .
\label{ampl}
\ee
Since we have found a second invariant contributing to the process,
we may wonder whether an appropriate linear combination of the
two invariants can reproduce the result of eq.~(\ref{amp}).
Indeed, it is immediate to check that, with the special choice 
$\alpha=-4$, the combination ${\cal L}_{eff}' + \delta {\cal L}_{eff}'$ 
reproduces the scattering amplitudes obtained from ${\cal L}_{eff}$.

As a first comment on the interpretation of our results,
we would like to stress that there is no reason to believe 
that the result of eq.~(\ref{leff}) is more fundamental than the 
standard result of eq.~(\ref{leffter}). The important fact to 
realize is that, since two independent invariants can be 
constructed, both of which contribute to the effective four-fermion 
coupling under consideration, there is an ambiguity in the effective 
theory description, parametrized by the coefficient $\alpha$ in 
eq.~(\ref{deltal}). At the level of the linear realization, this 
ambiguity is contained in the coefficients of higher-derivative 
operators, which are not included in the standard K\"ahler 
formulation of eq.~(\ref{expan}). Notice also that the new term 
(\ref{deltal}) scales with $F$ exactly as the term (\ref{leffter}),
which provides the coupling with $T_{\nu\mu}$. They both contain 
two derivatives and give rise to amplitudes with the same energy 
behaviour. Therefore, in the low-energy expansion of an underlying
fundamental theory, they are on equal footing. Moreover, the new 
supersymmetric invariant (\ref{invtwo}) gives rise only to terms 
containing at least two goldstinos, without modifying the free 
matter fermion lagrangian. 

Also, our results may admit a geometrical interpretation\footnote{We
thank S.~Ferrara for discussions and suggestions on this point.}.
Using the equations of motion and some Fierz identities, we can 
rewrite the contribution (\ref{deltal}) to the effective lagrangian 
as
\be
\delta {\cal L}_{eff}' = {\alpha \over 8 F^2} ( i \, \epsilon^{\mu 
\nu \rho \lambda} - \eta^{\mu \nu} \eta^{\rho \lambda}) 
[(\partial_{\nu} \psis) \sigma_{\rho} (\partial_{\mu} \psisb)] 
(\psie \sigma_{\lambda} \psieb) = {\alpha \over 8 F^2} 
(S^{\lambda} + T^{\lambda}) (\psie \sigma_{\lambda} \psieb) \, ,
\ee
where
\be
S^\lambda \equiv  i \, \epsilon^{\mu \nu \rho \lambda} 
(\partial_{\nu} \psis) \sigma_{\rho} (\partial_{\mu} \psisb) 
\, ,
\;\;\;\;\;
T^{\lambda} \equiv - \eta^{\mu \nu} 
(\partial_{\nu} \psis) \sigma^{\lambda} (\partial_{\mu} \psisb) 
\, ,
\ee
which suggests a possible coupling of the matter current
to a non-trivial torsion term for the goldstino manifold.

\vspace{1cm}
{\bf 5.} Are (\ref{invone}) and (\ref{invtwo}) the only independent 
invariants that contribute to the effective four-fermion coupling 
under consideration, or are there others? To answer this question,
we look for all the local supersymmetric operators that respect 
the $U(1)$ global symmetry associated with matter conservation,
and contribute to physical amplitudes with two goldstinos and two 
matter fermions that grow at most as $s^2$. Such operators have
dimension $d \le 4$, where the counting takes into account an overall
factor $1/F^2$, necessarily associated with the two goldstinos. We
do not consider operators with $d>4$ because the corresponding amplitudes 
are suppressed by further powers of energy. Since we will use the
superfields as building blocks, we recall that the matter superfield 
$E$ has $d=3/2$. For the goldstino, it is convenient to consider the 
rescaled superfield $\Lambda/\sqrt{2} F$, which has $d= -1/2$. In 
this way, the goldstino field $\psis$ always appears in the combination
$(\psis/\sqrt{2} F)$. Throughout this section we will use units such 
that $\sqrt{2} F=1$: the appropriate powers of $F$ can be recovered 
at the end, simply by counting the goldstino fields. Finally, the 
integration measure $d^4\theta$ has $d=2$, and an additional unit 
is associated with each explicit space-time derivative acting on the 
superfields. 

The lowest-dimensional operator containing two matter-fermion 
and two goldstino component fields is a $d=2$ four-fermion 
term of the kind $\psie\psis~\psieb\psisb/F^2$. Is this
allowed by supersymmetry? In terms of superfields, all 
the operators considered here contain precisely one matter
superfield $E$ and one conjugate matter superfield $\Eb$. In 
the absence of explicit space-time derivatives, the $d=2$ 
invariants require six goldstino superfields. Such operators 
vanish identically because of the Grassmann algebra, which allows no 
more than four goldstino superfields. For each explicit space-time 
derivative, two additional goldstino superfields are needed to keep 
the overall dimension constant, and the previous argument still 
applies. Therefore no local $d=2$ invariant is allowed by 
supersymmetry.

Moving to $d=3$, the only independent operator without
explicit space-time derivatives and (Pauli) $\sigma$-matrices
is $E\Lambda~\Eb\Lb~\Lambda^2$, up to an overall hermitean 
conjugation. However, this operator vanishes because of the 
Grassmann algebra. The result is unchanged if different Lorentz 
structures are considered, with any number of $\sigma$-matrices
and $\elc$ tensors inserted. Adding explicit 
space-time derivatives requires the inclusion of additional 
goldstino superfields, and the Grassmann algebra forces the 
corresponding operators to vanish. No $d=3$ invariant is 
permitted \footnote{Of course, by releasing the requirement of 
matter conservation or by adding additional matter superfields, 
$d=3$ invariants are allowed. They contain mass terms for the 
matter particles.}.    
 
We are left with the $d=4$ invariants. First, we consider the 
case of no explicit space-time derivatives. If $\sigma$-matrices
are also excluded, then the only possibility is the new invariant
$E\Lambda~\Eb\Lb$ of eq.~(\ref{invtwo}). Moreover, it is not 
difficult to see that, thanks to well-known properties of the 
$\sigma$-matrices, expressions involving an arbitrary number of 
$\sigma$'s and $\elc$ tensors always reduce to the invariant 
of eq.~(\ref{invtwo}).

When one space-time derivative is added, the independent invariants
containing only one $\sigma$ are, up to integration by parts and 
hermitean conjugation: 
\bea
S_1&=&(\dmd\Lambda)\smu\Lb~E\Lambda~\Eb\Lb \, , \nn\\
S_2&=&\Lambda\smu\Lb~E(\dmd\Lambda)~\Eb\Lb \, , \nn\\
S_3&=&(\dmd\Lambda)\smu\Eb~E\Lambda~\Lb^2 \, , \nn\\
S_4&=&\Lambda\smu\Eb~E(\dmd\Lambda)~\Lb^2 \, , \nn\\
S_5&=&\Lambda\smu\Eb~E\Lambda~\Lb(\dmd\Lb) \, , \nn\\
S_6&=&E\smu\Eb~\Lambda(\dmd\Lambda)~\Lb^2 \, , \nn\\
S_7&=&\Lambda\smu\Lb~\Lambda(\dmd E)~\Eb\Lb \, , \nn\\
S_8&=&\Lambda\smu\Eb~\Lambda(\dmd E)~\Lb^2 \, , \nn\\
S_9&=&(\dmd E)\smu\Lb~\Lambda^2~\Eb\Lb \, , \nn\\
S_{10}&=&(\dmd E)\smu\Eb~\Lambda^2~\Lb^2  \, .
\label{dim4}
\eea
The invariants $S_1,\ldots,S_6$ do not produce terms without goldstino
fields. We have explicitly evaluated the terms containing two matter 
fermions and two goldstinos, making use of integration by parts and of 
the equations of motion. The terms generated by $S_5$ and $S_6$ vanish. 
Those produced by $S_1$ and $S_3$ coincide, up to overall factors, with 
the operator of eq.~(\ref{deltal}). The terms coming from $S_2$ and 
$S_4$ are proportional to $(\psie\dmd\psis)(\psieb\dmu\psisb)$. The 
contributions of this four-fermion interaction to the helicity amplitudes 
for $\psie\psieb\to\psis\psis$ are however identical, up to overall 
factors, to those induced by the operator (\ref{deltal}). Therefore, 
the inclusion of the invariants $S_1,\ldots,S_6$ merely amounts to a 
redefinition of the parameter $\alpha$ in the amplitudes of 
eq.~(\ref{ampl}).

The invariants $S_7,\ldots,S_{10}$ give rise also to a term
proportional to the matter-fermion kinetic term in the lagrangian.
In particular, $S_{10}$ is the invariant that occurs for a massless 
fermion according to the prescription of refs.~\cite{nl,nlbis}, and 
that was already discussed in the previous section 
[see eq.~(\ref{invone})]. We have explicitly expanded 
the invariants $S_7,\ldots,S_9$ 
up to terms containing two goldstinos. Then we have evaluated, for 
each invariant, the contributions to the helicity amplitudes for the 
process $\psie\psieb\to\psis\psis$. Once the normalization of the 
kinetic term for the matter fermion is properly taken into account, 
such contributions are exactly the same as those originated from the 
invariant $S_{10}$, despite the occurrence, in the intermediate steps 
of the computations, of new four-fermion operators. Therefore, any 
combination of $S_7,\ldots,S_{10}$, such that the matter kinetic term
in the lagrangian is canonically normalized, gives rise to the physical 
amplitudes given in eq.~(\ref{ampter}), with no free parameters. This 
exhausts the case of one space-time derivative and one $\sigma$-matrix. 
All the invariants obtained by adding $\sigma$-matrices and $\elc$ 
tensors can be reduced to the invariants $S_1,\ldots,S_{10}$ by using 
properties of the $\sigma$-matrices. 

The next case involves two space-time derivatives acting on the 
superfields. The independent invariants with no $\sigma$'s are, 
up to integration by parts and hermitean conjugation:
\bea
S_{11}&=&E(\dmd\Lambda)~\Eb(\dmu\Lb)~\Lambda^2~\Lb^2 \, , \nn\\
S_{12}&=&E(\dmd\Lambda)~\Eb\Lb~\Lambda^2~\Lb(\dmu\Lb)\, , \nn\\
S_{13}&=&E\Lambda~\Eb\Lb~\Lambda(\dmd\Lambda)~\Lb(\dmu\Lb) \, .
\eea
They produce an interaction of the type $(\psie\dmd\psis)(\psieb\dmu
\psisb)$, as in the case of the invariants $S_2,S_4$. As we have 
seen, this does not affect the parametrization of the physical amplitudes 
provided by eq.~(\ref{ampl}). New invariants can be obtained by adding two 
$\sigma$-matrices. We have checked that the corresponding physical 
amplitudes are still given by eq.~(\ref{ampl}). More $\sigma$'s and 
$\elc$ tensors do not generate independent invariants.

Finally, having more than two derivatives requires more than six
goldstino superfields and the Grassmann algebra does not allow
to build non-vanishing combinations.

In conclusion, assuming matter conservation, the most general amplitudes 
for processes involving two goldstinos $\grav$ and two massless matter 
fermions $f$ can be parametrized in terms of only two supersymmetric 
invariants. The first one, eq.~(\ref{invone}), is normalized by the 
requirement of providing a canonical kinetic energy for the matter 
system. The second one, eq.~(\ref{invtwo}), brings a free parameter 
$\alpha$ in the expression of the amplitudes. No additional invariant 
is required, at least when only two goldstinos are present. This 
restricts the form of the helicity amplitudes. For instance, the
general amplitudes for the process $f \ov{f} \to \grav \grav$ are
just the sum of eqs.~(\ref{ampter}) and (\ref{ampl}), 
\be
\label{agen}
a_{GEN}(L,R,L,R) = {1 \over F^2} \left( t u 
- {\alpha \over 4} s u \right) \, ,
\;\;\;\;\;
a_{GEN}(L,R,R,L) = {1 \over F^2} \left( - t u 
+ {\alpha \over 4} s t \right) \, ,
\ee
where $(s,t,u)$ are the usual Mandelstam variables $[t=-(s/2)
(1-\cos \theta),u=-(s/2) (1+\cos \theta)]$, and the corresponding 
total cross-section is
\be
\label{sigmagen}
\sigma_{GEN} (f {\bar f} \to \grav \grav) = 
{(8 + 10 \alpha + 5 \alpha^2) s^3 \over 3840 \pi F^4} \, .
\ee
Notice that the cross-section (\ref{sigmagen}) is minimized
for $\alpha=-1$, with $\sigma_{min} = s^3 / (1280 \pi F^4)$.

\vspace{1cm}
{\bf 6.} We conclude with some remarks on the interpretation, 
the possible extensions and the phenomenological implications 
of our results.

It would be interesting to see how our results can 
be interpreted within the framework of supersymmetric current 
algebra, which was successfully used for the first derivations 
of supersymmetric low-energy theorems \cite{dwf}. We see a 
suggestive analogy with the textbook case of pion-nucleon 
scattering (see, e.g., section 19.5 of \cite{sw}), where the 
effective lagrangian consists of two independent terms, one 
completely controlled by the broken $SU(2) \times SU(2)$ 
symmetry and the other one containing the axial coupling $g_A$ 
as an arbitrary parameter.

It would be also interesting to generalize our framework by 
including gauge interactions, and make contact with the recent 
results of \cite{lp}. At the level of local four-fermion
operators, the arguments of the previous section are not
affected by the presence of gauge interactions\footnote{In 
particular, our proof implies that there are no $d=2$ local 
supersymmetric operators contributing to $e^+ e^- \to 
\grav \grav$ in the limit of vanishing electron mass.
If present, these operators would be characterized by a 
dimensionful coupling $M^2$, where $M$ is an independent mass 
scale, possibly arising from the underlying fundamental theory.}.
However, non-local four-fermion operators can in principle 
be generated by photon exchange, and this considerably 
complicates the discussion. We leave this to future 
investigations \cite{bfz4}. Since the process $e^+ e^- \to \grav 
\grav$ may be used to extract a lower bound on the gravitino 
mass from supernova cooling (for recent discussions, see
\cite{bfz3,lp,sn}), we expect a further clarification of 
this important phenomenological issue.

When extended to observable processes and realistic models, our 
results have other important phenomenological implications.
Consider for example the reaction $f \ov{f} \rightarrow
\grav \grav \gamma$, which probably gives the best 
signature of a very light gravitino at high-energy 
colliders, if all the other supersymmetric particles
are above threshold. Also in this case, the explicit integration
of the heavy superpartners gives results \cite{bfz4} that 
differ from those obtained \cite{nach} from the non-linear 
realization of \cite{nlbis}.  In our opinion, it would be
important to provide our experimental colleagues with a general 
framework to search for a superlight gravitino in a model-independent 
way, and we hope to develop this point soon.

\newpage

\end{document}